\documentclass{ar-1col-S2O}
\usepackage[numbers]{natbib}
\usepackage{url}
\usepackage{multirow}
\jname{Prepared for Annu. Rev. Nucl. Part. Sci.}
\jvol{AA}
\jyear{2026}

\begin{document}

\markboth{Koshio et al.}{SK upgrade with Gd}

\title{Upgrade of Super-Kamiokande with Gadolinium}

\author{Yusuke Koshio,$^1$ Masayuki Nakahata,$^2$ Hiroyuki Sekiya,$^2$ and Mark R. Vagins$^3$
\affil{$^1$Department of Physics, Okayama University, Okayama, Okayama 700-8530, Japan}
\affil{$^2$Kamioka Observatory, Institute for Cosmic Ray Research, University of Tokyo, Kamioka, Gifu 506-1205, Japan}
\affil{$^3$Kavli Institute for the Physics and
Mathematics of the Universe (WPI), The University of Tokyo Institutes for Advanced Study,
University of Tokyo, Kashiwa, Chiba 277-8583, Japan}}

\begin{abstract}
Super-Kamiokande [SK] was upgraded through the addition of gadolinium sulfate to its ultrapure water, initiating the SK-Gd program. This development enables efficient neutron tagging via the large capture cross section of gadolinium, greatly improving the identification of inverse beta decay events, the primary channel for detecting the diffuse supernova neutrino background [DSNB]. The upgrade also enhances sensitivity to galactic and pre-supernova neutrinos, as well as atmospheric neutrino interactions. To realize this capability, extensive work was performed, including the construction and operation of the EGADS demonstrator, the refurbishment of the SK tank, the development of radiopure gadolinium production methods, and the validation of the loading and uniformity of gadolinium in solution. Early SK-Gd operation has demonstrated high neutron-tagging efficiency, reduced backgrounds, and world-leading limits on the DSNB flux. With these advances, SK-Gd now stands at the threshold of discovering the DSNB and opens a wide range of new opportunities in astrophysics and neutrino physics.
\end{abstract}

\maketitle

\tableofcontents

\newcommand{\GdSOw}{$\rm Gd_2(\rm SO_4)_3\cdot \rm 8H_2O$}
\newcommand{\GdSO}{$\rm Gd_2(\rm SO_4)_3$}
\newcommand{\GdOx}{$\rm Gd_2 \rm O_3$}

\section{INTRODUCTION}
Light water Cherenkov [WC] detectors such as Kamiokande,
IMB, and Super-Kamiokande [SK; also known as Super-K], have been utilized for
many years to search for nucleon decays and to study neutrino
properties, interactions, and sources. Water -- which serves as 
both a target/source for particle interactions/nucleon decays and
as an active medium to convert the subsequent passage of 
energetic charged particles into Cherenkov light -- is readily available and inexpensive,
and can be made very transparent for efficient transmission of light. 
This light can then be collected by an encircling array of 
photomultiplier tubes [PMTs], which continuously observe the 
enclosed water volume for flashes of light.

\begin{figure}[htb]
\centering\includegraphics[width=0.9\linewidth]{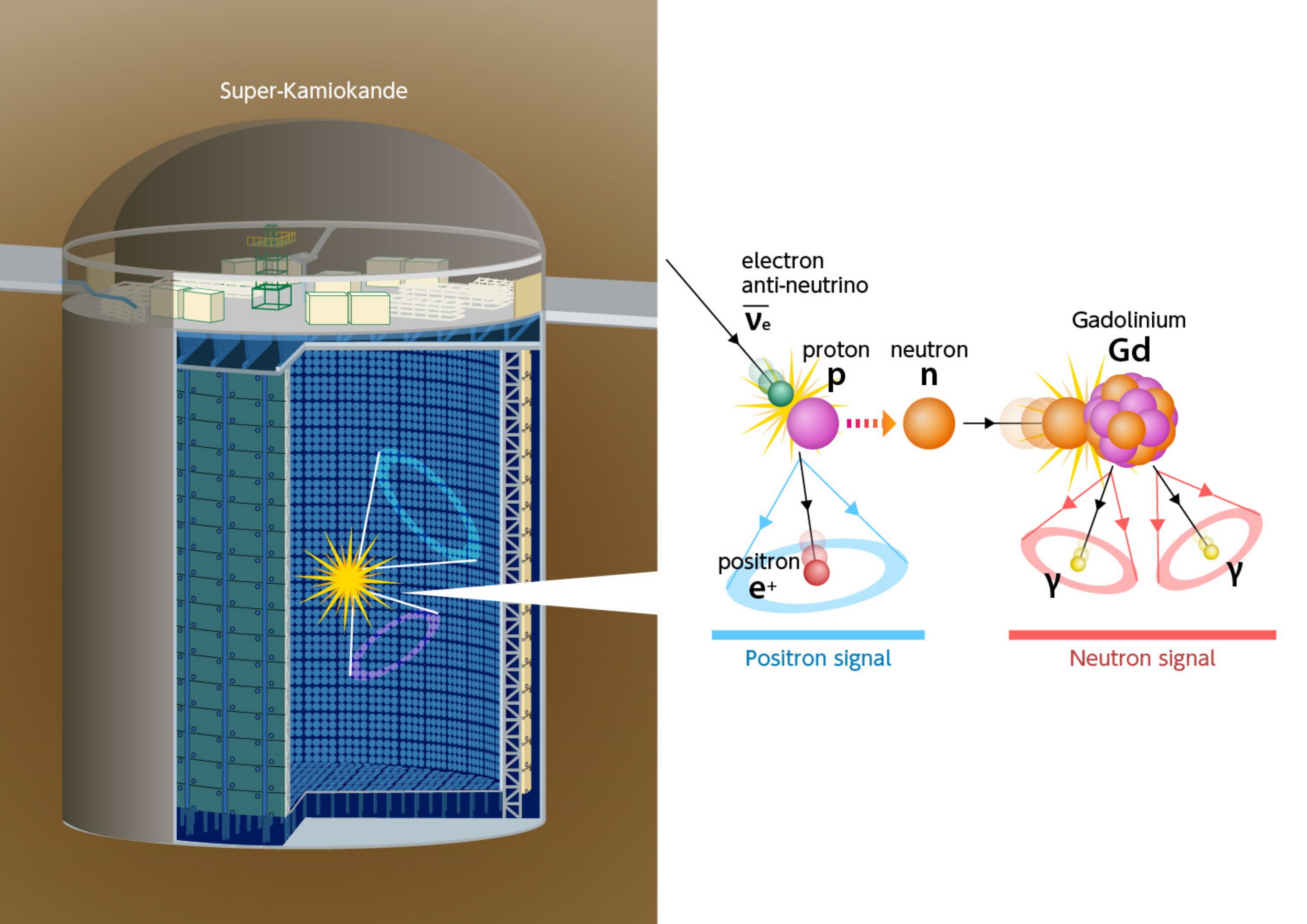}
\caption{Depiction of an inverse beta reaction in the Gd-loaded Super-Kamiokande detector.}
\label{fig:SKGd-image}
\end{figure}

While many important measurements have been made with these WC 
detectors, a major drawback has been their very inefficient 
response to the presence of thermal neutrons. Now, an affordable 
and effective technique has been developed and put into practice 
at Super-Kamiokande to overcome this situation via the addition to 
water of an optically transparent, low background, low reactivity 
solute with a large neutron cross section and energetic gamma 
daughters: gadolinium sulfate octahydrate (\GdSOw).  See  Figure~\ref{fig:SKGd-image}.

As discussed in Section 2, naturally occurring gadolinium [Gd] has 
the highest thermal neutron capture cross section of any stable 
element, with no need for isotopic enrichment, while Gd's  
efficient tagging of thermal neutrons enhances a variety of 
physics topics.  

The idea of loading the famous Super-Kamiokande neutrino and nucleon detector with tons of a water soluble gadolinium compound, a concept which would initially be called \textit{GADZOOKS!} [\underline{G}adolinium \underline{A}ntineutrino 
\underline{D}etector \underline{Z}ealously \underline{O}utperforming 
\underline{O}ld \underline{K}amiokande, 
\underline{S}uper\underline{!}]~\cite{beacom:2004} and then later 
officially put into practice as 
\textit{SK-Gd},  
was first introduced at the November 2002 SK Collaboration meeting. 

After investigating the use of gadolinium chloride 
and gadolinium nitrate as our Gd-containing solute, we settled on 
gadolinium sulfate as the best possible compound for use in Super-K due to its combination 
of high solubility and high transparency coupled with low reactivity 
with detector components. 
The year 2009 saw
the funding and construction of EGADS [\underline{E}valuating 
\underline{G}adolinium's \underline{A}ction on \underline{D}etector 
\underline{S}ystems], a dedicated, large-scale 
test facility for gadolinium R\&D~\cite{egads:2020}.  This new underground laboratory 
was built in the Mozumi mine explicitly to demonstrate that the 
loading of Super-Kamiokande would be safe and effective (see Section 3).

After a major in-tank refurbishment in 2018/9 (see Section 4) 
and a campaign to produce tons of the world's most radio-pure gadolinium 
sulfate~\cite{Hosokawa:2023} (see Section 5), in 2020, some eighteen years after the idea was 
first presented, 
the first gadolinium was successfully added to Super-Kamiokande~\cite{1stload:2022}, 
with a second loading to increase the concentration 
of dissolved Gd$^{3+}$ ions taking place in 2022~\cite{2ndload:2024} (see Section 6).

Finally, the initial results of SK's ongoing Gd-loaded phase will 
be shown (in Section 7), and future prospects will be discussed 
(in Section 8).

\section{PHYSICS BENEFITS OF GADOLINIUM IN SUPER-KAMIOKANDE}
\subsection{Neutron Capture in Water}\label{sec:neutroncap}
In the energy region of supernova neutrinos, the dominant detection reaction in water is inverse beta decay [IBD] : $\bar{\nu_e} + p \rightarrow e^+ + n$~\cite{STRUMIA200342}.
In this reaction, the emitted positron's Cherenkov light is detected as a prompt signal (the eventual annihilation of the positron is not visible in water), followed by gamma rays generated by the capture of the neutron by a nucleus, which are detected as a delayed signal.  
This detection method is called the delayed coincidence method and is much more effective in reducing background events compared to measuring only the prompt signal.
Furthermore, the original neutrino energy can be easily reconstructed from the prompt positron energy using the following equation: $E_{\nu}\sim E_{e^+}+m_n-m_p$, where $m_n$ and $m_p$ are the neutron and proton mass, respectively.

In pure water, the emitted neutrons thermalize and are captured by protons, with the newly formed deuterons each emitting a single 2.2~MeV gamma ray which can then be measured as delayed signals.
Since the fourth phase of SK [SK-IV] started in 2008, new electronics equipment has been introduced~\cite{Yamada:2010}, making it possible to measure delayed signals.
However, in order to achieve the background reduction required for the DSNB analysis mentioned below, the neutron detection efficiency had to be set relatively low (to about 20\%). But the situation changes with the introduction of gadolinium.
When a neutron is captured by gadolinium, it emits a cascade of gamma rays with a total energy of about 8~MeV, as depicted in   Figure~\ref{fig:SKGd-image}. 
This is higher than the energy threshold in SK, and thus the detection efficiency improves drastically even under the same trigger conditions as in the pure-water case.
Furthermore, gadolinium has an exceptionally high neutron capture cross section compared to all other elements; it will capture about 75\% of thermal neutrons with a modest concentration of only 0.03\% Gd by mass dissolved in the water.

\subsection{Diffuse Supernova Neutrino Background}
Core-collapse supernovae are powerful explosions that occur in the final stages of the evolution of massive stars, during which roughly 
99\% of their enormous release of gravitational energy is carried away by neutrinos.
Since the beginning of the universe, there have been $\mathcal{O}(10^{18})$ core-collapse supernovae, and the neutrinos released in these explosions fill the current universe as background radiation. This is called the diffuse supernova neutrino background [DSNB].
DSNB observations are expected to provide valuable insights into astrophysics, such as past star formation rates, the frequency of supernova explosions, and the formation of black holes which cannot be seen by optical observations~\cite{beacom:2010, lunardini:2016, Suliga:2022, Ando:2023}.
Since the 1980s, DSNB searches have been conducted using various detectors; however, before the start of SK-Gd no nonzero DSNB flux had been observed.
Super-Kamiokande currently has the most stringent constraints on this flux~\cite{Harada_2023, AbeSN:2021, bays:2012}, but the expected event rate of DSNB interactions even in the huge SK detector is quite low - at most a few per year - and they are buried under background events, making observation extremely difficult.

The most significant background source in the DSNB search is caused by muon spallation, and the delayed coincidence method is very effective in removing these events.
At Super-Kamiokande [SK], located 1,000 meters underground, the flux of cosmic ray muons can be reduced to 1/100,000th of that at sea level; however, there are still about two muons arriving per second.
When muons enter the SK detector, they spall oxygen nuclei in the water, producing various radioactive isotopes.
Events that occur when these radioactive isotopes decay become background events and are called muon spallation events.
These events can be reduced using the correlations of position and time with the muon track, yielding a reduction efficiency of 99\% (i.e., $10^{-2}$). 
However, to perform an effective DSNB search, especially in the lower energy region, a total reduction of $10^{-6}$ is required. The delayed coincidence method 
with efficient neutron tagging provided by gadolinium is used to gain the additional needed reduction factor of  $10^{-4}$.
As mentioned above, the introduction of gadolinium has proven to be very effective in reducing these background events while retaining the signal events.

Other background events are caused by atmospheric neutrinos.
When primary cosmic rays, mainly consisting of protons, collide with atomic nuclei in the Earth's atmosphere, various hadrons, such as pions and kaons, are produced.
When these hadrons decay, atmospheric neutrinos are generated.
The energies of atmospheric neutrinos range from $\sim$100 MeV to PeV, peaking at approximately 100~MeV.
The particles produced by atmospheric neutrinos interacting in the detector have a wide energy range, and when these particles have reconstructed energies in the DSNB region, they become background events. 
Again, the use of neutron signals helps to distinguish atmospheric neutrino events, as they can produce single, multiple, or zero neutrons, whereas DSNB anti-electron neutrino signals always produce one neutron.
Furthermore, atmospheric neutrinos producing one neutron can be background events, but by studying such events in the energy region higher than the DSNB (so-called sideband region), these interactions' total contribution in the region of interest can be estimated.

Thus, the introduction of gadolinium into Super-Kamiokande has made it possible to capture neutrons with higher efficiency than pure water, improving the sensitivity of the DSNB search.
Of course, a further understanding of the remaining background events is necessary, but the current observational limit is approaching the DSNB flux prediction, and we can expect the first observations to occur in the near future.
These will be discussed in Sections 7 and 8.

\subsection{Galactic Supernova Neutrinos}
As mentioned above, many supernovae have occurred to date, but supernova neutrinos have only been observed once.
In 1987, neutrinos from a supernova explosion that occurred in the Large Magellanic Cloud, 50 kpc from Earth, were observed at Kamiokande-II, IMB, and Baksan, with 11, 8, and 5 neutrino events, respectively\cite{Hirata:1987, Bionta:1987, Alexeyev:1987}.
This observation proved that the basic mechanism of gravitational core collapse that occurs in the final stages of star evolution is correctly hypothesized.
Unfortunately, the rate of supernovae close enough to detect neutrinos is estimated at only a few per century~\cite{Adams:2013}, so the chance of one occurring within the next decade is low, around 10-30\%.
However, if such a supernova occurs and a large number of neutrino events are observed, it would be a great advance in our understanding of core-collapse supernovae.
Therefore, neutrino observatories around the world are currently preparing for the next nearby supernova to occur.

If a supernova occurs at the center of our galaxy, Super-Kamiokande will detect thousands of neutrinos.
Here, the dominant neutrino interaction is the IBD described above, but if a large number of neutrinos are detected, we can also expect detection through other interactions.  These include, for example,  
elastic scattering [ES] : $\nu + e^- \rightarrow \nu + e^-$; charged current interaction on oxygen : $\nu_e + \rm{^{16}O} \rightarrow e^- + \rm{^{16}F}$, $\bar{\nu_e} + \rm{^{16}O} \rightarrow e^+ + \rm{^{16}N}$; and neutral current interaction on oxygen : $\nu + \rm{^{16}O} \rightarrow \nu + \rm{^{16}O^*} \rightarrow \nu + \rm{^{15}O/^{15}N} + \rm{n/p} + \gamma$ ~\cite{Nakanishi:2024}.
In particular, the direction of electrons produced by ES is strongly correlated with the direction of the original neutrinos, and therefore ES events can be used by Super-Kamiokande to determine the location of the supernova in the sky.
Here, it is extremely important to determine the direction to the supernova as quickly as possible.
Neutrinos are made \emph{before} the huge increase in brightness of the dying star, and thus these neutrinos arrive at Earth earlier -- anywhere from minutes to hours, depending on the stellar environment and type of progenitor star -- than the supernova's electromagnetic signals. 
Consequently, if the sky location of an ongoing explosion obtained from neutrino observations can be transmitted quickly enough to the astronomical community, then it should be possible for them to observe the beginning of the shock breakout of a core-collapse supernova.

An important point in determining the direction of a supernova is how efficiently the ES signal (about 5\% of all events) can be extracted from data containing a large number of nearly isotropic IBD signals.
Again, the key here is neutron detection: IBD emits a neutron, but ES does not; therefore, the accuracy of the determination of the direction to a supernova can be improved by reconstructing the direction using only neutron-free events.
In particular, the use of gadolinium in SK-Gd is expected to improve the accuracy of determining the supernova direction compared to the previous pure water phase because of the improved neutron detection efficiency.
In fact, it has been shown that the pointing circle on the sky in SK-Gd will have a radius of about 4~degrees (at 68\% C.L.) for a supernova occurring 10~kiloparsecs away from Earth, a distance which would include more than 50\% of the stars in our galaxy~\cite{Kashiwagi:2024}.
Identification of supernovae with this accuracy would be useful for follow-up observations with wide-field optical telescopes.


\subsection{Pre-supernova Neutrinos}
As mentioned in the previous section, many neutrinos are expected to be seen by SK-Gd when a core-collapse supernova occurs within our galactic neighborhood, but if the explosion is relatively close to Earth ($<$1 kpc) neutrinos could also be observed just before core collapse.
Stars during the final stage of their lives are supported by nuclear fusion of heavy nuclei such as silicon, and the cooling mechanism is mainly by neutrino emission via thermal and nuclear processes. 
These are called pre-supernova neutrinos~\cite{Odrzywolek:2004}.
Although pre-supernova neutrinos have not yet been found, their observation is very important because this silicon burning phase is thought to last for several days before the core collapse; detecting these neutrinos opens up the possibility of creating a supernova alert.
It also would serve to guard against the unfortunate case of SK being down for planned calibration or maintenance just when a nearby star explodes.
However, because the flux of these pre-supernova neutrinos is low, the possibility of their measurement is limited to very large, very nearby stars such as Betelgeuse.
In fact, Super-Kamiokande has achieved highly efficient neutron capture in SK-Gd and significantly reduced background events, making the detection of neutrinos before a supernova explosion a realistic goal.
Since 2021, SK-Gd has been monitoring pre-supernova neutrinos.
The sensitivity of the system strongly depends on the supernova model and neutrino mass ordering, but in the case of Betelgeuse, for example, it should be  possible to issue an alarm between 2 and 12 hours before its core collapse at a significance level corresponding to a false alarm rate of no more than 1 per century~\cite{Machado:2022}.
In addition, joint observations with the KamLAND experiment began in May 2023.
By integrating the complementary characteristics of the two detectors, the combined alarm system improves its sensitivity to pre-supernova neutrinos.
For example, in the most optimistic model, the warning time is 1.5 hours longer than with SK-Gd alone.
The combined alert system also reduces the dead time for pre-supernova neutrino detection and facilitates continuous monitoring even if one of the detectors is temporarily offline~\cite{Abe:2024}.\footnote{As is currently the case 
for KamLAND at the time of this writing.  That detector is temporarily offline for a major upgrade that started in 2024 and is expected to 
be completed in 2027.}

\subsection{Others} 
Super-Kamiokande is a multipurpose detector, and the increased neutron detection efficiency from gadolinium loading has benefits for a variety of physics purposes.
The advantages of each are described below.

\subsubsection{Reactor Neutrinos}
One of the physics opportunities of SK-Gd is the observation of reactor neutrinos.
Since the spectrum of reactor neutrinos is soft, the detection is overwhelmed by radioactive backgrounds and hopeless if we use only the prompt signals of IBD.
However, if a delayed neutron signal can be detected, the background events can be drastically reduced, opening the possibility of observing reactor neutrinos.
Nevertheless, it is necessary to realize a lower energy threshold for the prompt signal as compared to the DSNB search.
This is a long-baseline reactor neutrino observation, famous for measuring neutrino oscillations for $\theta_{12}$ and $\Delta m^2_{21}$ in the KamLAND experiment~\cite{Eguchi:2003}.
In SK-Gd, reactor neutrinos are expected to be observed on the order of 100 per year from the currently limited cohort of operating Japanese reactors.

\subsubsection{Atmospheric Neutrinos}\label{sec:atmnu-bene}
In GeV-scale neutrino observations, improving the neutron detection efficiency using Gd is expected to improve the ability to distinguish between neutrinos and antineutrinos.
This is because antineutrinos are more likely to produce neutrons in the primary interaction than neutrinos, and the number of neutrons detected after neutrino interactions is expected to reflect this tendency.
However, the number of neutrons produced by secondary interactions within the nucleus increases with energy, and thus the final state neutron multiplicity is much larger than would be predicted simply by considering only primary interactions.
Therefore, understanding hadron-nucleus interactions inside the nucleus and in the detector is crucial for this purpose.
For atmospheric neutrinos, the ability to distinguish between neutrinos and antineutrinos improves the sensitivity to neutrino mass ordering by taking advantage of the different oscillation patterns of their neutrinos as they pass through the Earth's core and mantle~\cite{Wester:2024}.

\subsubsection{Accelerator Neutrinos}
T2K is a long-baseline neutrino experiment in which accelerator neutrinos generated at J-PARC in Tokai, Ibaraki, Japan, are observed at SK, approximately 300 km away.
This experiment is expected to also improve T2K's measurement sensitivity by improving the ability to distinguish between neutrinos and antineutrinos, as mentioned in the previous section~\cite{Abe:2025}.
Increased neutron detection efficiency is particularly important for antineutrino appearance analysis, as contamination of neutrino events is currently expected to be relatively large.
Preliminary studies have shown that this contamination is reduced from 30\% to 13\% by gadolinium loading.
In this study, we have confirmed that, although still very preliminary, the introduction of additional $\bar{\nu}$-enhanced and $\nu$-enhanced samples slightly improves the sensitivity in the analysis of CP violation in the lepton sector.

\subsubsection{Proton Decay}
The main background for proton decay originates from atmospheric neutrinos.
In general, fewer neutrons are emitted after proton decay than in the background events caused by atmospheric neutrinos; therefore, efficient neutron detection can eliminate more background, improving proton decay sensitivity.
For example, in the standard $p \rightarrow e^+ + \pi^0$ mode, the probability that one or more free neutrons are present in the final state is approximately 7\%.
On the other hand, most background events resulting from atmospheric neutrino interactions have one or more neutrons in the final state, and neutron tagging with gadolinium is estimated to reduce the atmospheric neutrino background by 83\%.
Thus, adding neutron information significantly improves the sensitivity of proton decay searches.
It should be noted that the benefit of neutron tagging with gadolinium is even greater when only one candidate event is observed.
For example, after 10 years of observation, the expected background in pure water is 0.58 events, while it is reduced to 0.098 events with gadolinium.
The Poisson probabilities of observing one or more events with these expected backgrounds are 44\% and 9\%, respectively.
Thus, even with a single candidate event, only when gadolinium is used can we claim to have a proton decay signal with a confidence level of 90\% or higher.

\section{EGADS DEMONSTRATOR}
\begin{figure}[htb]
\centering\includegraphics[width=0.9\linewidth]{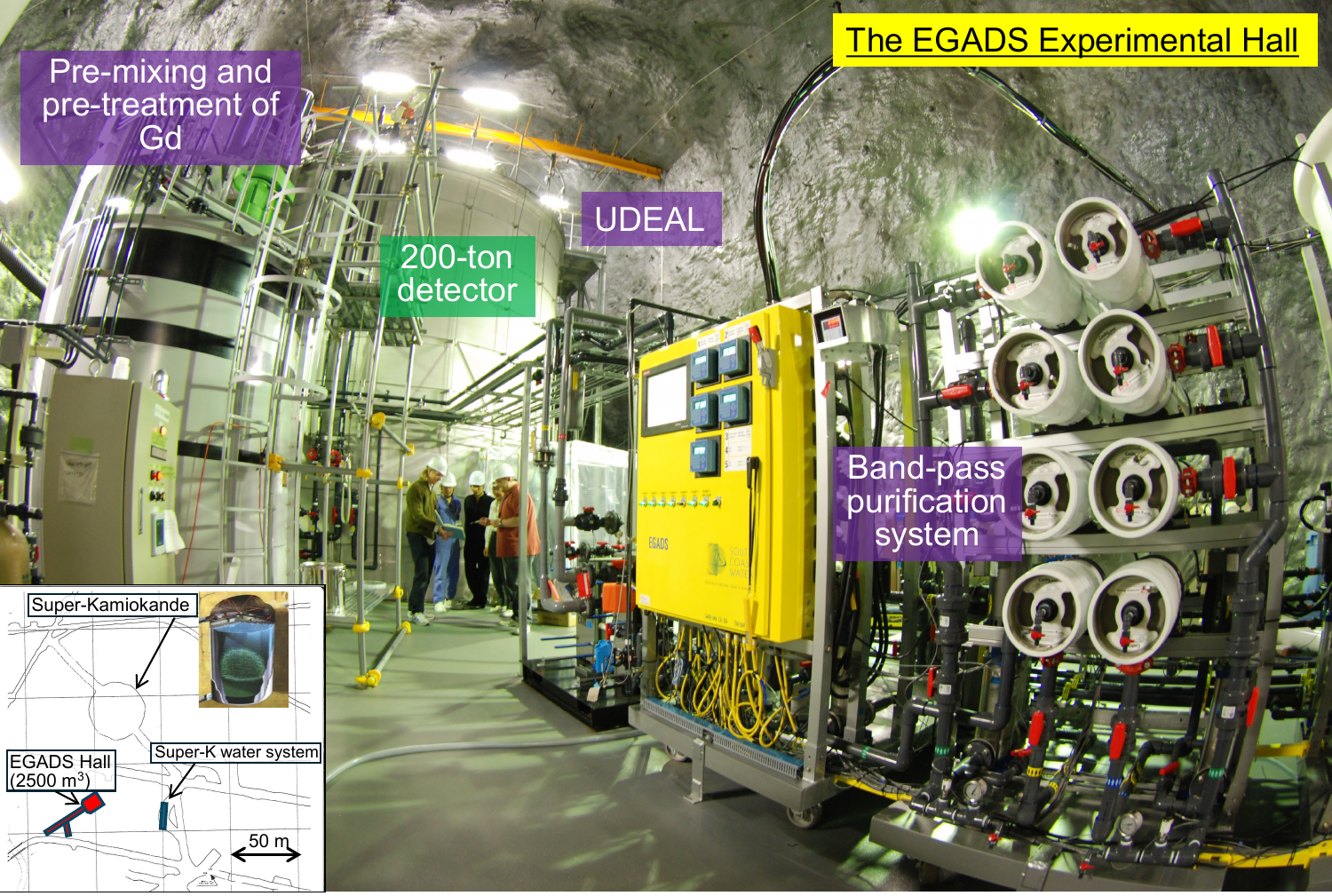}
\caption{The EGADS experimental hall in the Kamioka mine.}
\label{fig:egads}
\end{figure}

In order to demonstrate the safety and effectiveness of the Gd-in-water  approach, a dedicated gadolinium test facility and technology demonstrator was constructed underground in the Kamioka mine near Super-Kamiokande, as can be seen 
in Figure~\ref{fig:egads}. This large-scale R\&D project is called EGADS 
[\underline{E}valuating 
\underline{G}adolinium's \underline{A}ction on \underline{D}etector 
\underline{S}ystems].

The EGADS facility includes a 200 ton scale model of SK complete with
240 photomultiplier tubes (227 50-cm SK-style tubes plus an additional
13 prototype tubes that were evaluated for use in the Hyper-Kamiokande
project\footnote{Hyper-Kamiokande [HK] is the next-generation successor to Super-Kamiokande, a similar but much larger WC detector containing eight times SK's fiducial volume~\cite{HK:2021}. HK is currently under construction in a neighboring mountain in Kamioka and is expected to begin data-taking in 2028.}), and two different selective water filtration systems.  There 
is also a 15-ton pre-mixing and pre-treatment tank where the \GdSOw~is 
dissolved and cleaned prior to injection into the main
200-ton tank, and a custom-built water transparency measuring system 
called UDEAL [\underline{U}nderground \underline{D}evice  
\underline{E}valuating \underline{A}ttenuation \underline{L}ength]. 

Following 18 months of construction, EGADS first became operational in
2011, initially running with pure water. Various adjustments and tuning
of the equipment followed, and by 2013, once it had been demonstrated
that the filtered pure water in EGADS was equal in quality to that in
SK, 400 kilograms of gadolinium sulfate – a world’s record at the time – 
was dissolved in the 200 ton main tank, allowing large-scale studies of
gadolinium filtration and transparency to begin.  The two 
independent EGADS selective
filtration systems, one membrane-based (called the ``band-pass") and 
the other resin-based (called the ``fast recirculation"), 
have since achieved gadolinium retention rates of 
$>99.97$\% per pass, while simultaneously cleaning unwanted 
impurities from the water and indefinietly maintaining - in the presence of gadolinium - water transparency comparable to that in SK's pure water phases: an unprecedented result and a major 
technical achievement.  In addition, the dissolved \GdSOw~was 
shown to not damage the SK-style detector components, even 
after years of exposure.

These EGADS studies ultimately demonstrated that putting gadolinium in SK should work exactly as predicted~\cite{beacom:2004}. As a result, the SK-Gd project was approved, leading to the SK tank being drained and refurbished in 2018/9 in preparation for the loading of gadolinium 
as discussed in the next section.

Many more details regarding this essential testbed and its measurements can be found in our dedicated EGADS journal article~\cite{egads:2020}.

\section{TANK REFURBISHMENT FOR GD-LOADING}

Since the beginning of Super-K’s operation in 1996, there had been a water leak
from the Super-K tank, ranging from 1 to 4 tons per day, depending on the water
level in the surrounding rock.
That amount of leakage was acceptable as long as the system was running with pure
water because purified fresh water was periodically added to the tank.
In the case of Gd-loaded water, we had to consider environmental issues.
There exist no environmental standards for gadolinium anywhere in
the world, including in Japan.  Consequently, Gd and its compounds 
are not classified as regulated chemicals by Japan's Water 
Pollution Prevention Act. 
However, release of Gd-loaded water to the environment should be avoided  
because the hazard levels for gadolinium compounds are not yet well 
established. 
That is one of the main reasons why we performed refurbishment work
from May 31, 2018, to January 29, 2019.
The main tasks were as follows:
\begin{itemize}
    \item Fix a small water leak in the Super-K tank.
    \item  Clean any rust and other dirt accumulated in the detector.
    \item  Install additional water piping to increase the total water flow for
  improved water purification and to enable
better control of the flow direction in the tank.
    \item Replace the PMTs that have
failed (a few hundred out of 13,000) since the previous
in-tank refurbishment in 2006.
\end{itemize}

\subsection{The Leak Fix}
\begin{figure}
\includegraphics[width=0.8\textwidth]{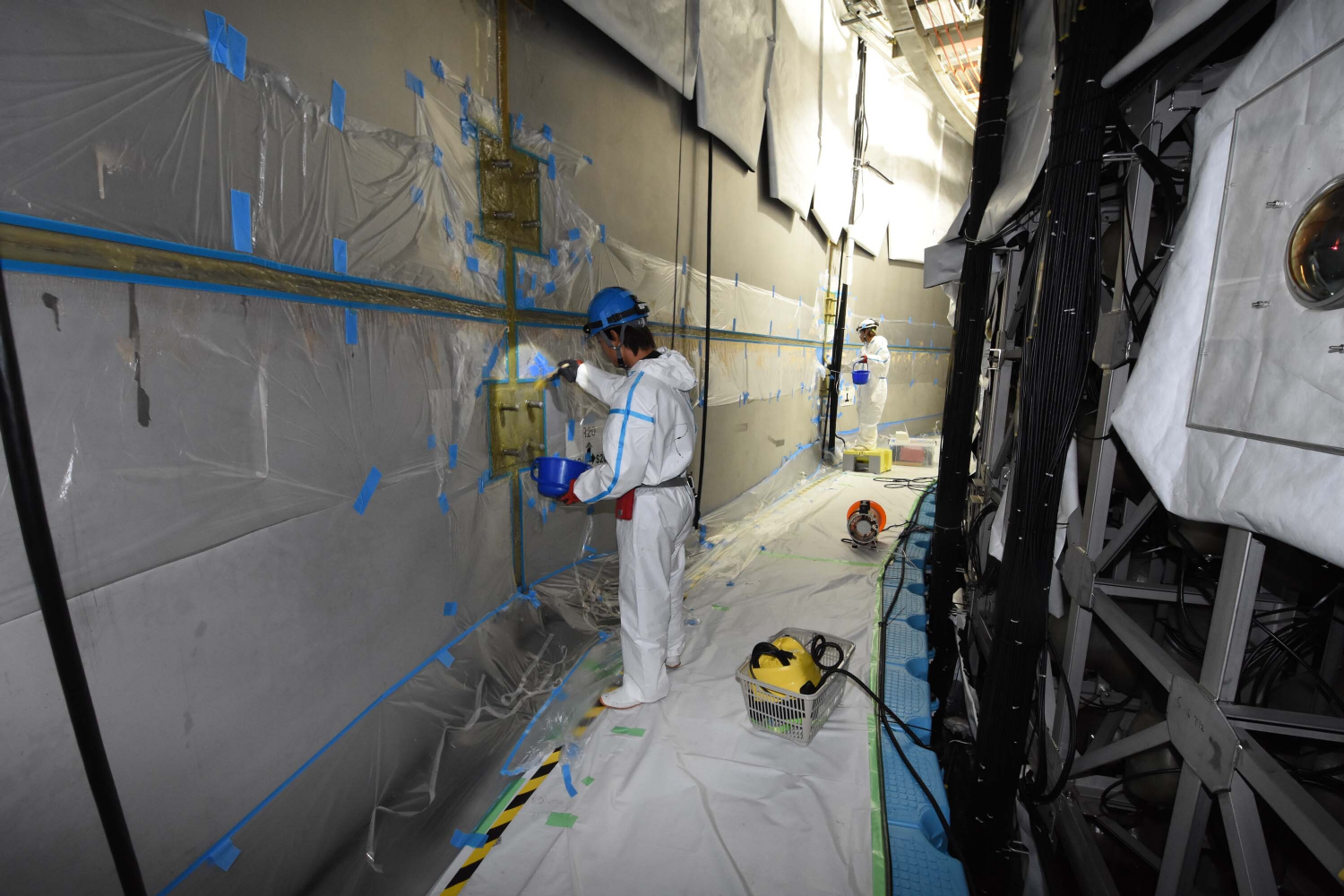}
 \caption{Sealant work on the SK tank wall conducted from a 
floating floor}
  \label{fig:seal}
\end{figure}

During this refurbishment effort, we drained two meters of
water from the tank every three days and worked to seal potential water
leakage points on the outer wall and replace those failed PMTs
that we could reach from a floating floor on the water surface.
The Super-K tank is 40~meters in diameter and 42~m tall, and contains
50,000 tons of pure water.
Its sides form an icosagon (a 20-sided polygon) comprised of 400 pieces of 4~millimeter-thick stainless
steel plates 6~m wide and 2~m tall, which are able to withstand external stretching forces with
the support of backup concrete and rock.
Similarly, the bottom of the tank is comprised of 120 pieces of 3~mm-thick stainless steel plates.
The basic principle of the leak fix was to paint sealant on all
welded places, including the welding lines between stainless steel plates and around all supporting structures that protrude from the tank wall.
The total length of the welding lines amounts to about 6200~m.
Figure~\ref{fig:seal} shows the actual sealant work.
Workers put thick tape along both sides of the welding line to hold the
sealant material on the line.
They used trowels and paint brushes to spread the sealant material over the weld
with the same thickness as the surrounding tape.
A double sealant layer was required to allow for unexpected gaps or
pinholes in one of the two layers.
The sealant material (product name: “MineGuard C6”) was newly developed in 
collaboration with Hodogaya Construction Product Co., Ltd. 
This material demonstrates sufficient adhesive strength on stainless steel 
plates, even in Gd-loaded water. 
It also has the necessary elasticity to accommodate the expansion and 
contraction of welded areas on stainless steel plates and maintains a low U/Th 
concentration to keep radon emissions as low as the current Super-K level. 
Several improvements have been made to the original MineGuard product. 
For enhanced hydrolysis resistance, the base resin was changed from 
polyurethane to polyurea. 
To achieve a sufficiently low U/Th concentration, the viscosity-controlling 
agent was replaced with fumed silica (CAB-O-SIL M5 from Cabot Corporation) 
instead of calcium-based materials. 
For the primer, we used the A10 product from Shin-Etsu Chemical Co., Ltd.

After filling the tank with pure water, a water leakage measurement
was conducted.
The method monitors the water level to 0.1 mm precision while water
circulation is stopped.
We could not see any decrease in the water level and
concluded that the SK water leakage has been reduced to less
than 1/200th of the rate during the period before the tank upgrade work.

\subsection{Cleaning and Rust Removal}
The dirty walls and structures inside the Super-K tank were cleaned
during the refurbishment period.
In particular, the top and outermost structures were covered by a
heavy residue of exhaust gas and dust left over from when we
originally constructed the detector in 1991-1996.
We also removed most potential sources of rust during
the cleaning work. 
These areas had been acceptable during the pure water phase of SK
in terms of water transparency but the dirt would be easily released
to Gd-loaded water and degrade the water transparency.
We maintained the excellent quality of sealant attachment and
removed a large amount of radioactivity through the dirt removal;
according to a germanium detector measurement, at least 100 Bq of
radium was removed during the cleanup.
The radon gas emitted from this radium is comparable to that
arising from the glass of the PMTs.

\subsection{Water Piping Upgrade}
The SK water purification system was originally designed with a supply
rate of 30~m$^3$/h and circulation rate of 60~m$^3$/h, and the piping inside the tank was built without separate lines for the inner
detector [ID] and the outer detector [OD].
During the 2018/2019 tank open work we upgraded the water piping to
increase the circulation flow rate to 120~m$^3$/h.
In addition, we improved the water flow in
the tank by separating the pipes of the inner volume and the outer
volume and the pipes on the side, top, and bottom of the outer volume
so that all flow rates could be adjusted independently.
The effective detector volume for the solar neutrino analysis had
been limited by the inflow of radon background, but this plumbing
improvement makes it possible to control and optimize the water
flow in the tank~\cite{1stload:2022}.

\subsection{PMT Exchange and Other Works}
During this tank open period, dead PMTs were also exchanged
in parallel with sealing work.
In total, 136 ID PMTs and 224 OD PMTs were exchanged.
For ID PMT replacement, new PMTs under development for
Hyper-Kamiokande [HK] were used.
In addition to the recovery of SK performance, another
purpose of this exchange was to check the performance and
long-term stability of HK PMTs in the expected operating environment.
Another important task carried out during the refurbishment period 
was related to the calibration system inside the tank and studies
for more precise detector understanding, such as the replacement
of laser injectors and optical fibers, installation of a
new light injection system, precise magnetic field mapping,
and recording all PMTs' dynode directions.
All the white Tyvek sheets covering the tank wall were removed
for the sealant work, with new Tyvek sheets re-installed afterward.

\section{PURE GD PRODUCTION AND SCREENING}

The \GdSOw\ used for SK-Gd had to meet stringent radiopurity standards to ensure that any added background remained subdominant to those from existing SK solar neutrino and DSNB measurements~\cite{Hosokawa:2023}. Specifically, the total increase in background rates, even after reaching the target concentration of 0.1\% Gd in SK, was required to be no more than a factor of two.

To satisfy these requirements, detailed specifications were developed through simulations and experimental studies~\cite{egads:2020,pablo:2017phD,Ito:2020ptep}, and an extensive R\&D program was conducted to develop chemical processing and purification methods capable of producing ultra-clean \GdSOw\ powder~\cite{Hosokawa:2023}.

\subsection{Radiopurity Requirements}

The primary contaminants of concern were uranium and thorium isotopes, as their decay products can either mimic IBD signals (relevant for DSNB detection) or contribute to backgrounds in low-energy solar neutrino measurements. The requirements are determined by limiting contributions from:

\begin{itemize}
  \item $^{238}$U spontaneous fission [SF], which mimics IBD signals and affects DSNB measurements.
  \item $\beta$-decay backgrounds overlapping the solar neutrino energy range from the $^{232}$Th and $^{238}$U chains.
  \item Neutron production via ($\alpha$,n) reactions from $^{235}$U.
\end{itemize}

The criteria for each relevant isotope, along with the associated physics targets and corresponding parts per billion (ppb) limits, are summarized in Table~\ref{tab:RIsummary}.



\subsection{Purification Method}

To meet the strict radio-purity requirements for SK-Gd, a multi-step purification method was developed to produce ultra-pure \GdSOw~starting from gadolinium oxide (\GdOx). The raw material \GdOx, with $>$99.99\% purity (Gd$_2$O$_3$/Total Rare Earth Oxide), was selected with special care to minimize thorium and uranium concentrations.

\subsubsection{Dissolution and Solvent Extraction}

\GdOx\ was dissolved in hydrochloric acid to form a gadolinium chloride solution with a Gd concentration of $\sim$305~g/L. After removing insoluble impurities, the solution was subjected to solvent extraction using 2-ethylhexyl 2-ethylhexylphosphonate [PC-88A] diluted in isoparaffin (20:80 ratio). At a controlled pH of 1.0--1.3, radioactive impurities such as Th and U were extracted into the organic phase while Gd remained in the aqueous phase. This extraction step was repeated to ensure high purification efficiency.

\subsubsection{Neutralization and Sulfation}

Further removal of Th and Ce was achieved through neutralization steps. Thorium was precipitated as hydroxide at pH~4.8, and cerium was oxidized and precipitated as cerium oxide using 0.3\% hydrogen peroxide. Finally, 98\% sulfuric acid was added to precipitate \GdSOw, which was then washed until pH~4 and dried. The final product was confirmed to be the octahydrate form via X-ray diffraction and thermogravimetric analysis.

\subsubsection{Additional Purification Process for Radium Reduction} \label{sec:new_purification}

As described in Section~\ref{dissolving}, the introduction of \GdSOw, into the SK tank was carried out in two stages: an initial 13-ton loading, followed by a second 26-ton loading. For the latter, an additional purification step was implemented before the solvent extraction to reduce radioactive impurities further, specifically $^{226}$Ra. The new purification process is based on the difference in precipitation behavior between gadolinium and radium hydroxides. By adjusting the pH of an acidic aqueous solution containing gadolinium chloride to between 8 and 9, gadolinium selectively precipitates as Gd(OH)$_3$ while radium remains dissolved in the solution. In this process, ammonia was chosen as the base for pH control due to its minimal risk of introducing additional metal contaminants such as Na or K. A 25\% aqueous ammonia solution was added gradually to the Gd chloride solution, prepared by dissolving \GdOx, until a pH of 8.0 was reached. The resulting Gd(OH)$_3$ precipitate was then separated by filtration and washed until the chloride ion concentration of the filtrate was reduced below 1000~mg/L. This purified hydroxide was subsequently redissolved in a hydrochloric acid solution to form a new Gd feedstock for solvent extraction.

\subsection{Material Assay} \label{sec:screening}

Samples from each 500~kg batch of \GdSOw\ were assayed to determine whether the requirements for radioactive and fluorescent impurities are satisfied. Because the radioactive isotopes that contaminate \GdSOw\ may not be in secular equilibrium with their long-lived parents or daughters, high-purity germanium [HPGe] gamma spectrometry was employed to measure the activities of both the early and late parts of each decay chain relevant to SK-Gd physics sensitivities. In addition, inductively coupled plasma mass spectrometry [ICP-MS] is used to determine the concentrations of U and Th down to very low levels. The detailed assay results can be found in Refs~\cite{Hosokawa:2023,2ndload:2024}.

\subsubsection{Radioisotope Measurements Using HPGe Detectors}
\label{sec:Ge}

HPGe detectors located at the Boulby Underground Screening [BUGS] facility in the UK, the {\it Laboratorio Subterráneo de Canfranc} [LSC] in Spain, and the Kamioka Observatory in Japan~\cite{Ichimura:2023} were used to screen for radioactive contaminants. Most batches were measured at multiple laboratories to ensure consistency across sites. 

Due to the need to wait for the decay of background $\rm{^{222}Rn}$, each HPGe measurement in Kamioka takes approximately 20~days. The final two batches were delivered to the SK-Gd site about one week before dissolution, leaving insufficient time for $\rm{^{226}Ra}$ measurement after radon decay. To address this, a chemical separation method was used to evaluate the $\rm{^{226}Ra}$ concentrations in these two batches~\cite{Sakakieda:2023}. The measured concentrations of $^{226}$Ra were 0.84$\pm$0.05 and 0.23$\pm$0.02~mBq/kg, including procedural blanks (i.e., everything used in the measurement \emph{other} than the \GdSOw\ itself). Therefore, the intrinsic contamination levels are likely lower, and both batches were deemed acceptable. 

\subsubsection{ICP-MS for U, Th}

To measure uranium and thorium at the parts-per-trillion [ppt] level, a nitric acid aqueous solution of dissolved \GdSOw\ was passed through a well-washed chromatographic extraction resin that retains more than 90\% of the U and Th content. These elements were then eluted using a dilute nitric acid solution. Finally, the eluate was analyzed by ICP-MS to quantify trace amounts of U and Th without interference from gadolinium, which is reduced by a factor of approximately $10^4$ through this process~\cite{Ito:2017ptep}.

\subsection{Radioisotope Summary for Physics}
\label{sec:RIsummary}

Table~\ref{tab:RIsummary} summarizes the HPGe and ICP-MS measurement results for all the batches of \GdSOw~for the second Gd loading, and the combined results for both the first and the second Gd loadings.
The table also includes the required activity for each radioactive decay chain. The \textit{total budget} indicates the acceptable decay rate in SK-Gd 
for the case where 130 tons of \GdSOw\ are dissolved (corresponding to a 0.1\% Gd concentration).

The \textit{finite value} in the HPGe column represents the sum of all finite measured activities, with errors combined in quadrature. The \textit{upper limit} represents the conservative upper bound on the total activity, taking into account the sum of all 95\% confidence level upper limits.

The ICP-MS results show that contamination by $^{232}$Th and $^{238}$U is sufficiently low. The later parts of the decay chains, such as the $^{238}$U decay chain ($^{238}$U and $^{226}$Ra equivalent), were measured with HPGe detectors, and the contamination is sufficiently small. The $^{235}$U decay chain contamination ($^{235}$U and $^{227}$Ac equivalent) was also measured by HPGe detectors, and the 95\% upper limits for this decay chain are well below the total budget.

The $^{232}$Th decay chain, which includes $^{228}$Ra and $^{228}$Th, is required to be less than 6.5~Bq budget. However, due to insufficient sensitivity for this decay chain, it is difficult to confirm if the 6.5~Bq budget has been achieved. The total activities of $^{228}$Ra and $^{228}$Th are $<$18.6~Bq and $<$15.9~Bq, respectively.

\begin{table}[!ht]
    \centering
    \caption{Summary of HPGe and ICP-MS measurements for the second Gd-loading, the combined first and second Gd-loading, and total SK-Gd radioactivity budget assuming 0.1\% Gd-loading (130 tons of \GdSOw). The results for each radioactive chain are presented for the parent isotopes [RIs], early part [E], and late part [L]. HPGe assays provide estimates of the minimum and maximum total added radioactivity to SK. This table is taken from~\cite{2ndload:2024}.}

    \scalebox{0.75}[0.75]{
    \begin{tabular}{ccccccccccccccccc} \hline
         \multirow{5}{*}{Chain}       &    & &                               & & \multicolumn{4}{c}{2nd loading}              & & \multicolumn{4}{c}{1st + 2nd loading} \\ \cline{6-9} \cline{11-14}
                                      &    & \multicolumn{2}{c}{Requirement} & & \multicolumn{2}{c}{HPGe} & & ICP-MS        & & \multicolumn{2}{c}{HPGe}  & & ICP-MS \\ \cline{3-4} \cline{6-7} \cline{9-9} \cline{11-12} \cline{14-14}
                                      & Part of          & Specific & Total  & & Finite        & Upper    & & Total         & & Finite        & Upper    & & Total \\
                                      & Chain            & Activity & Budget & & Value         & Limit    & & (Bq)          & & Value         & Limit    & & (Bq)  \\
                                      &                  & (mBq/kg) & (Bq)   & & (Bq)          & (Bq)     & &               & & (Bq)          & (Bq)     & &       \\ \hline \hline
          \multirow{3}{*}{$^{238}$U}  & RI $^{238}$U     & $<$5     & 650    & & --            & --       & & 0.54$\pm$0.01 & & --            & --       & &  0.88$\pm$0.15   \\
                                      & E, $^{238}$U Eq. & $<$5     & 650    & & 0             & $<$183   & & --            & & 0             & $<$272   & &  --   \\
                                      & L,$^{226}$Ra Eq. & $<$0.5   & 65     & & 3.76$\pm$0.43 & $<$10    & & --            & & 4.0$\pm$0.4   & $<$15.6  & &  --   \\
          \hline
          \multirow{3}{*}{$^{232}$Th} & RI $^{232}$Th    & $<$0.05  & 6.5    & & --            & --       & & 0.21$\pm$0.01 & & --            & --       & &  0.46$\pm$0.07   \\
                                      & E, $^{228}$Ra Eq.& $<$0.05  & 6.5    & & 2.14$\pm$0.48 & $<$11    & & --            & & 5.4$\pm$0.6   & $<$19.7  & &  --   \\
                                      & L, $^{228}$Th Eq.& $<$0.05  & 6.5    & & 1.89$\pm$0.4  & $<$8     & & --            & & 5.6$\pm$0.5   & $<$17    & &  --   \\
          \hline
          \multirow{2}{*}{$^{235}$U}  & E, $^{235}$U Eq. & $<$30    & 3900   & & 0             & $<$22    & & --            & & 4.1$\pm$0.8   & $<$37    & &  --   \\
                                      & L, $^{227}$Ac Eq.& $<$30    & 3900   & & 0             & $<$23    & & --            & & 3.3$\pm$0.7   & $<$42    & &  --   \\ 
    \hline
    \end{tabular}
    }
    \label{tab:RIsummary}
\end{table}




\section{GD-LOADING AND DETECTOR UNIFORMITY}
\label{loading}
\subsection{Gd-loading system and water recirculation system}
\label{watersys}

The SK-Gd water system, newly designed for dissolving \GdSOw~into detector water, removes impurities and continuously purifies 50,000~tons of Gd-loaded water. This system preserves gadolinium (Gd$^{3+}$) and sulfate (SO$_4^{2-}$) ions while eliminating contaminants such as bacteria and particulates. Water was continuously circulated and purified throughout the Gd loading and operation periods.

\begin{figure}[htb!]
\centering\includegraphics[width=0.8\linewidth, trim = 150 0 150 0]{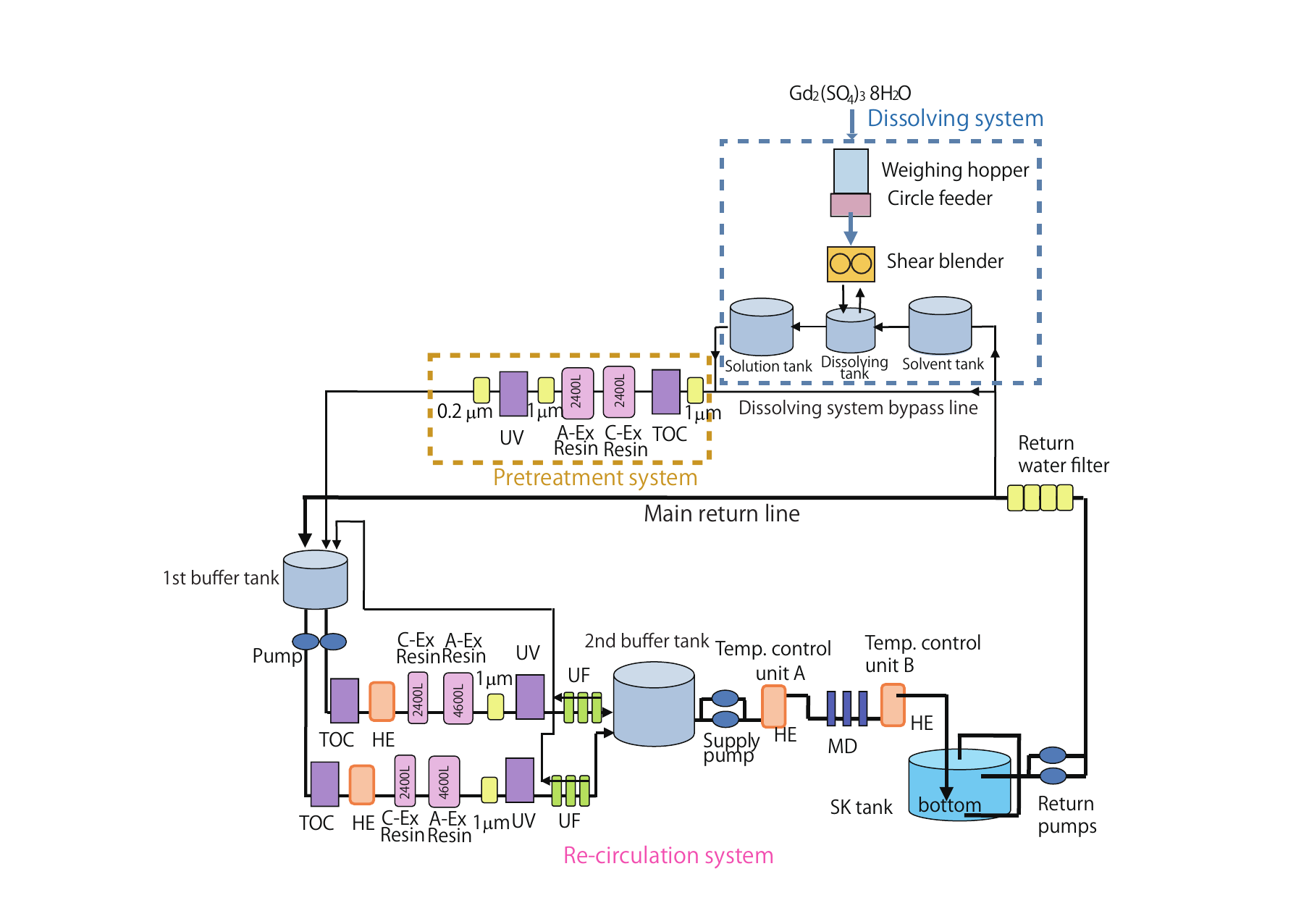}
\caption{Schematic diagram of SK-Gd water system. This figure is taken from Ref.~\cite{1stload:2022}. }
\label{fig:gd-water-sys}
\end{figure}

A schematic system diagram is shown in Fig.~\ref{fig:gd-water-sys}.

\subsubsection{Gd-dissolving System}

The Gd-dissolving system consists of a powder transportation section and a dissolution process. \GdSOw~powder is introduced into a weighing hopper, transferred via a circle feeder to the dissolving tank, and mixed with SK water using a shear mixer. 
The resulting solution is then sent to the solution tank before being pumped to the pretreatment system.


\subsubsection{Pretreatment System}

The pretreatment system removes impurities other than gadolinium and sulfate ions; this technology was developed and tested using the 200-ton EGADS detector~\cite{egads:2020}. After passing through a passive filter, the water undergoes ultraviolet [UV] irradiation to oxidize organic carbon, followed by ion-exchange resins that eliminate ionized contaminants such as uranium and radium. The system utilizes an anion exchange resin (AMBERJET 4400), conditioned with sulfate ions as the exchange group, to remove negatively charged impurities, particularly uranium compounds. Cation exchange resins (AMBERJET 1020 and IR120A) modified to contain gadolinium as the ion exchange group allow gadolinium to be retained while removing other metal ions. 

\subsubsection{Water Recirculation System}

The water recirculation system maintains transparency by continuously filtering Gd-loaded water. Two parallel lines allow circulation at 60~m$^3$/h per line, or 120~m$^3$/h in total. The filtration process includes ultrafiltration modules and ion exchange resins with a cation-to-anion ratio of $1:2$. Heat exchangers stabilize temperature, and a membrane degasifier removes dissolved radon. Water rejected by ultrafiltration [UF] modules is reprocessed to minimize waste, while losses from evaporation and degasification are replenished from a dedicated degasified ultrapure water refill system.

\subsection{Implementation and Validation of Gadolinium Loading}
\label{dissolving}
Gadolinium was introduced into the Super-Kamiokande detector in two sequential phases. The initial loading commenced on July 14, 2020, involving the dissolution of 13 tons of \GdSOw~\cite{1stload:2022}. This corresponded to a concentration of 0.026\% \GdSOw~by weight in 50,000 tons of ultrapure water, or equivalently 0.01\% by weight of elemental Gd, yielding a neutron capture efficiency of approximately 50\%.

To ensure uniform dispersion of gadolinium throughout the detector volume within a single full water circulation period (approximately 35 days), ultrapure water was extracted from the top of the tank, passed through the gadolinium sulfate dissolution system, and reintroduced at the bottom. This procedure facilitated rapid homogenization of the Gd concentration across the detector volume. The loading process proceeded stably and was completed on August 17, 2020.

\begin{figure}[htbp] 
\includegraphics[width=1.0\textwidth]{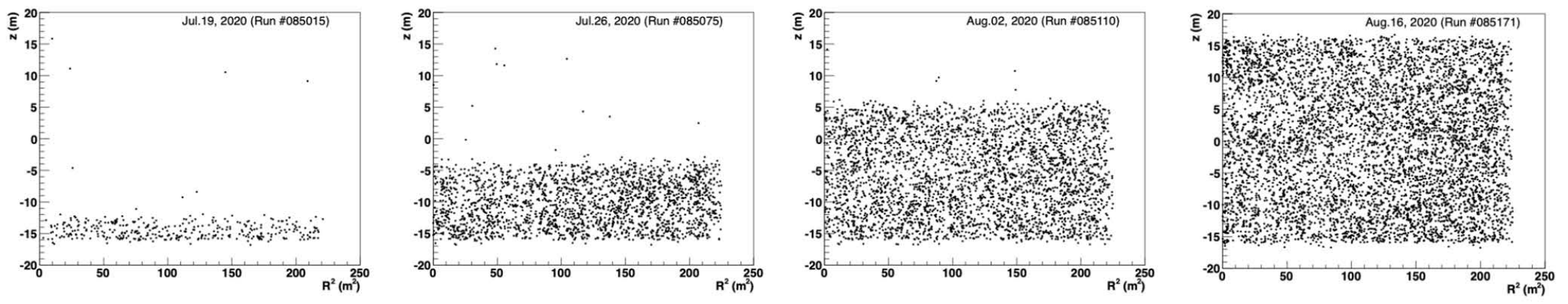} 
\caption{Time evolution of neutron capture events on gadolinium over 35 days during the first Gd loading. The horizontal axis represents the square of the radial position [m$^2$] in the cylindrical detector, and the vertical axis indicates the height [m].} 
\label{fig:1st_spa} 
\end{figure}

Figure~\ref{fig:1st_spa} illustrates the spatial distribution of neutron capture events on Gd, originating from muon-induced hadronic spallation during the first loading period. The daily evolution confirms the successful introduction of Gd from the bottom of the tank, as designed.

The Gd concentration was quantitatively assessed using an americium/beryllium [Am/Be] neutron source in conjunction with a bismuth germanate [BGO] scintillator. The scintillation light emitted by BGO crystals in response to gamma rays was used to simulate the expected signals from inverse beta decay events.

\begin{figure}[htbp] 
\includegraphics[width=0.6\textwidth]{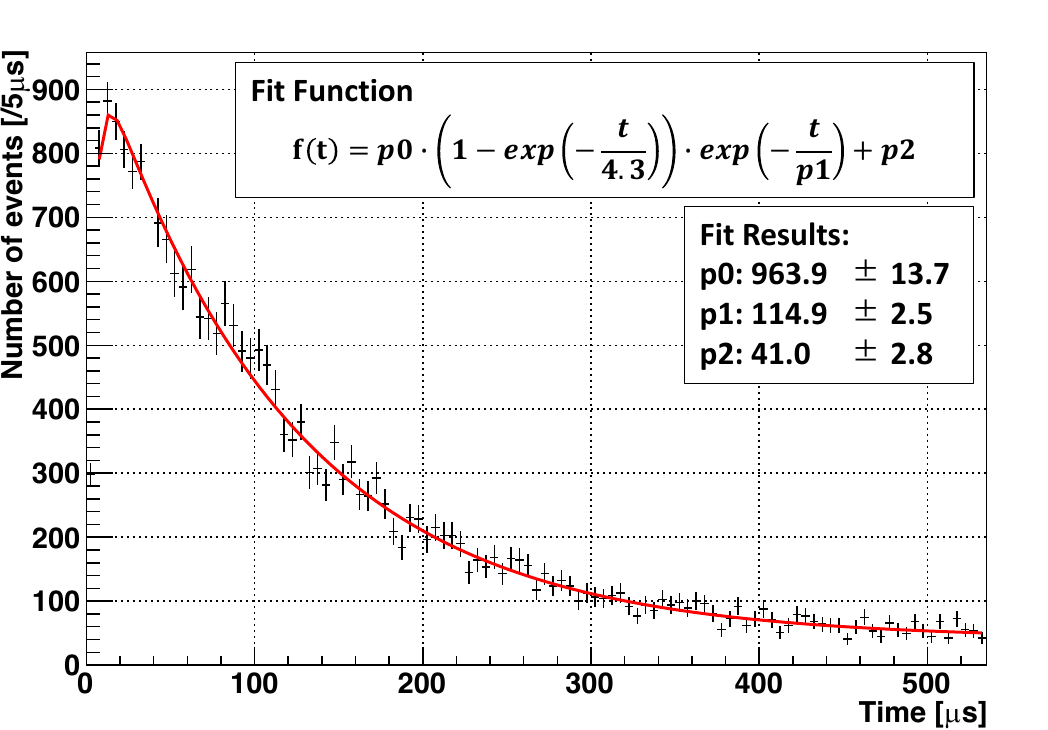} 
\caption{Distribution of time differences between scintillation events and neutron capture events on Gd in SK-VI, measured with Am/Be and BGO sources.} 
\label{fig:timediff} 
\end{figure}

Figure~\ref{fig:timediff} presents the distribution of time delays between scintillation and subsequent neutron capture events. The observed neutron capture time constant of approximately 110~$\mu$s is consistent with the expected value for a 0.01\% Gd concentration. Furthermore, spatial uniformity of the Gd concentration was verified by relocating the source to various positions within the tank. This phase of operations following the first Gd loading, characterized by a 0.01\% Gd concentration, was designated as SK-VI, the sixth official data-taking period of Super-Kamiokande.

\begin{figure}[htbp] 
\includegraphics[width=1.0\textwidth]{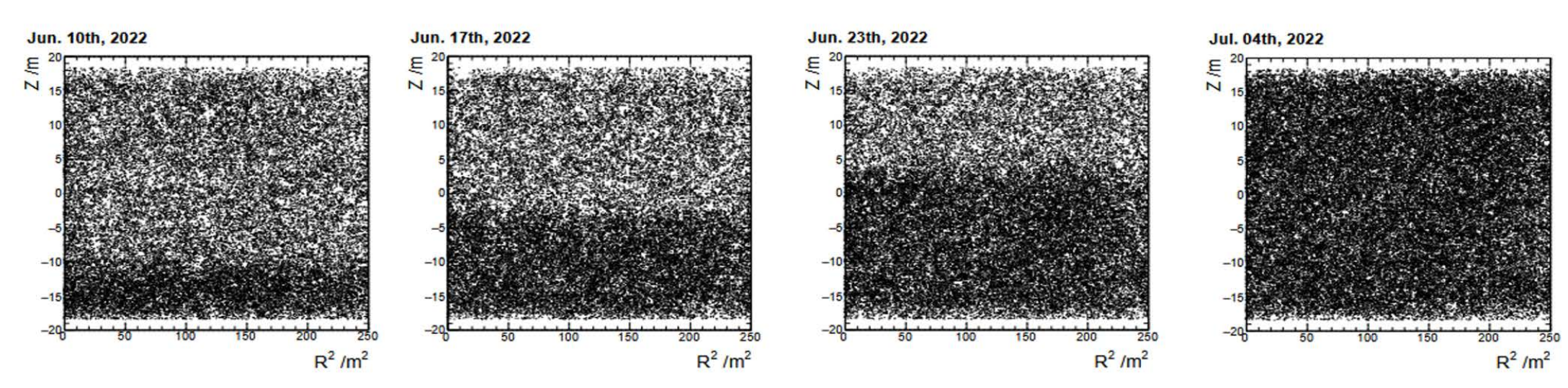} 
\caption{Time evolution of neutron capture events on gadolinium over 35 days during the second Gd loading phase.} \label{fig:2nd_spa} 
\end{figure}

The second phase of Gd loading was conducted from June 1 to July 5, 2022. In this phase, an additional 26 tons of \GdSOw~were dissolved, raising the overall Gd concentration to 0.03\% by mass and increasing the neutron capture efficiency to approximately 75\%~\cite{2ndload:2024}. As with the first loading, water containing Gd was introduced from the bottom of the tank over a single full circulation. The spatial uniformity of the Gd concentration following the second loading was confirmed via neutron capture event distributions, as shown in Figure~\ref{fig:2nd_spa}. This  operational phase following the second Gd loading was designated SK-VII.

\begin{figure}[htbp] 
\includegraphics[width=1.0\textwidth]{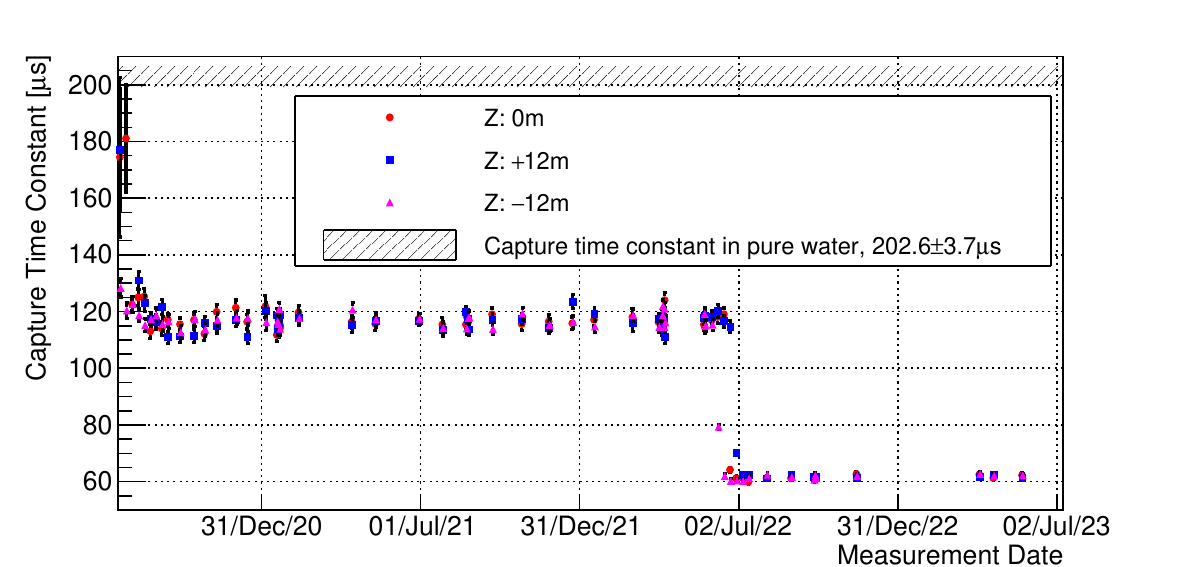} 
\caption{Time evolution of the neutron capture time constant in SK since July 2020, as measured using Am/Be neutron source data.} 
\label{fig:history} 
\end{figure}

Figure~\ref{fig:history} summarizes the periodic measurements of the Gd neutron capture time constant since the initial loading phase. The time constants, measured at various vertical positions within the detector, demonstrate consistent and stable values across both the SK-VI and SK-VII periods. The corresponding Gd concentrations inferred from these measurements are in agreement with the expected values based on the amount of \GdSOw~introduced, validating the long-term stability and uniformity of the system and confirming readiness for high-precision physics analyses.


\section{RESULTS OBTAINED IN SK-VI AND SK-VII}
\subsection{Search for DSNB}
We conducted a DSNB analysis on 552.2 and 404.0 days of data from the gadolinium-loaded SK-VI and SK-VII phases, respectively, and summarize the results in this section.  A separate paper detailing the Gd-enhanced DSNB analysis and results is in preparation~\cite{AbeDSNB:2025}.

Since the electronics upgrade in 2008~\cite{Yamada:2010} and for events with energies greater than about 6~MeV, in addition to the PMT hit signal from this ``prompt" event we can create a 535~$\mu$second-wide timing window and collect all hits occurring within this window.
We search for ``delayed" neutron capture events within this window.
Thanks to the electronics upgrade, a new event selection technique using the detection of the associated neutrons, called ``neutron tagging", has become available, drastically improving the sensitivity of the DSNB search.
The first analysis using this method was performed on SK-IV, the penultimate (and longest) pure water phase which lasted from 2008 to 2018~\cite{AbeSN:2021}.
In this phase, the delayed signal was a 2.2~MeV gamma ray emitted when a neutron was captured by a proton, but due to these gammas' low energy the pre-gadolinium neutron detection efficiency in SK was less than 20\%. 

This tagging efficiency has now been significantly increased in SK-Gd as described in Section~\ref{sec:neutroncap}.
Delayed neutron events are currently reconstructed by two independent machine learning methods: boosted decision trees [BDT] and neural networks [NN].
As described in detail elsewhere~\cite{AbeTag:2022}, hit clusters above a certain threshold within a given timing window are searched for, and feature variables for each machine learning method are calculated to produce an output score.
As explained in Section 2.2, the delayed coincidence method with Gd-enabled neutron tagging is required to reduce accidental coincidence events by $10^{-4}$, and the cut value of the machine learning score is selected so that the probability of accidental coincidence meets this condition.
For this cut condition, the efficiency of selecting one neutron is more than 60\%.

\begin{figure}[htb]
\centering\includegraphics[width=0.9\linewidth]{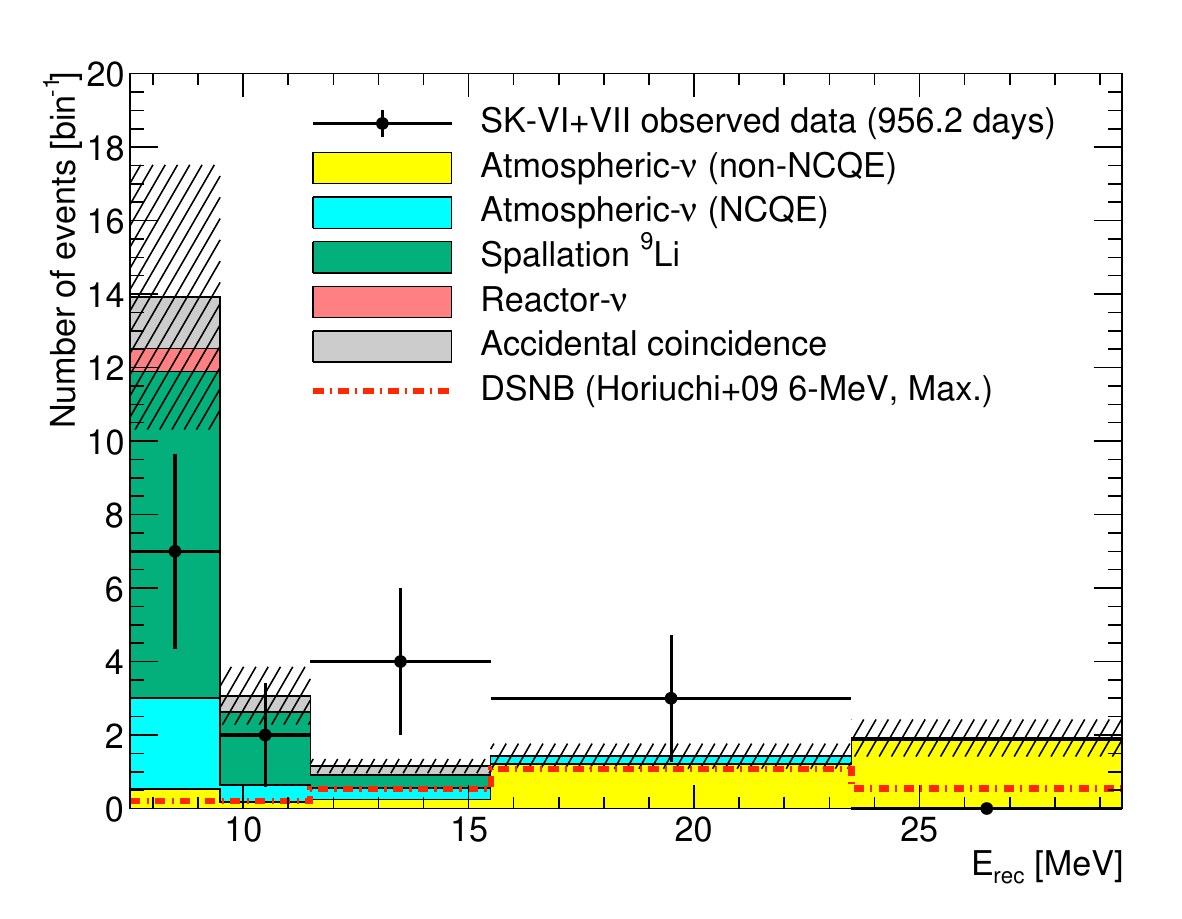}
\caption{Reconstructed positron equivalent kinetic energy distribution of the data and expected backgrounds. The error bars in data points show the statistical error. Here NN is used for neutron tagging method. Color-filled histograms show the expected backgrounds, which are stacked on top of each other. The hatched areas represent the total absolute background systematic uncertainty for each bin. For illustrative purposes, the 2009 DSNB expectation from Horiuchi et al.~\cite{Horiuchi:2009} is drawn separately from (not stacked on top of) the background histograms as a red-dashed line.}
\label{fig:dsnb_energy}
\end{figure}

The properties of the prompt events, such as vertex, direction, and energy, are reconstructed using the same algorithms as in the SK solar neutrino analysis~\cite{AbeSolar:2024}.
The method of selecting candidate DSNB events to reduce background has been described in detail in \cite{Harada_2023, AbeSN:2021, AbeDSNB:2025}.
These reductions have reduced the remaining background to a level approximately equivalent to the expected DSNB signal.
Figure~\ref{fig:dsnb_energy} shows the reconstructed energy distribution of the prompt events after all cuts are applied.
We concluded that there was no significant DSNB signal, although a slight excess was seen in some bins.

\begin{figure}[htb]
\centering\includegraphics[width=0.85\linewidth]{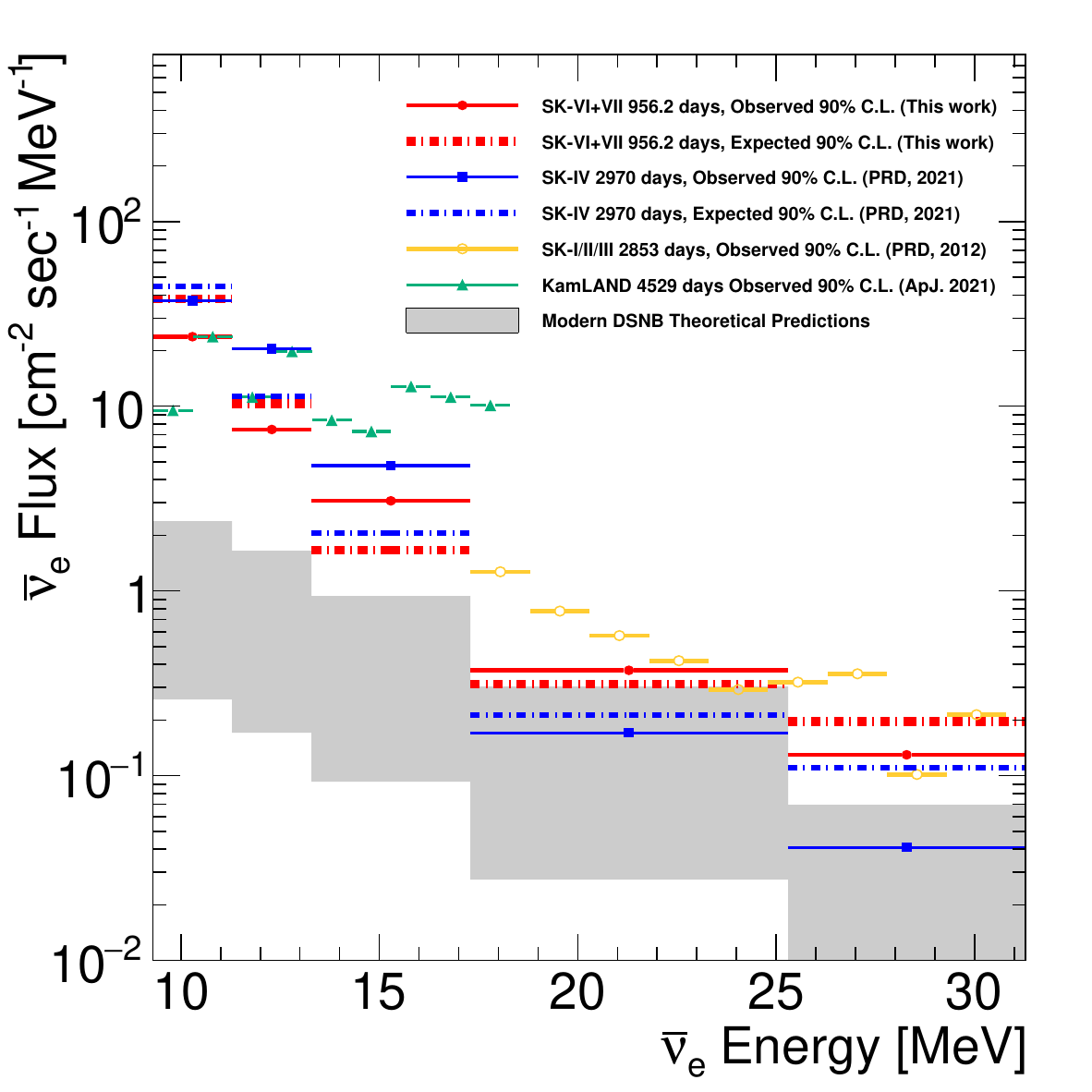}
\caption{90\% C.L. upper limits on the astrophysical
$\bar{\nu_e}$ flux per energy bin. The red lines show the observed upper limit (solid, circle point) and expected sensitivity (dot-dashed) for SK-VI+VII with Gd~\cite{AbeDSNB:2025}. Here NN is used as the  neutron tagging method. The blue lines show the observed upper limit (solid, square point) and expected sensitivity (dot-dashed) for SK-IV, the longest pure-water phase~\cite{AbeSN:2021}. The orange line displays the 90\% C.L. observed upper limit for the combined earlier pure-water phases, SK-I+II+III, without neutron tagging~\cite{bays:2012}. The green line with triangle points represents the 90\% C.L. observed upper limit placed by KamLAND~\cite{abe:2022b}. The gray-shaded regions represent the range of several theoretical flux expectations for the DSNB signal.
}
\label{fig:nuebar_limit}
\end{figure}

Since no statistically significant excess of events was observed above the expected background, we derived upper limits on the astrophysical electron antineutrino flux for each reconstructed energy bin.
For each energy bin, both the expected and observed 90\% confidence limits were calculated.
The resulting limits are shown in Figure~\ref{fig:nuebar_limit}.
Here, the expected limit is based on the number of events predicted from background alone, while the observed limit uses the actual number of detected events.
The sensitivity achieved above approximately 17~MeV is comparable to several theoretical predictions for DSNB.
In the energy range between about 13 and 17~MeV, the obtained limits approach the most optimistic DSNB models within a factor of two.
Relative to the previous Super-Kamiokande search using pure water~\cite{AbeSN:2021}, the new SK-Gd data provide improved sensitivity below 15.5 MeV owing to the significantly higher neutron-tagging efficiency and the reduced rate of accidental coincidences.
At higher energies, however, the pure-water results continue to offer the world’s best sensitivity for now, primarily due to smaller systematic uncertainties associated with atmospheric neutrino events and the larger exposure time of those datasets.

It is important to understand and reduce the remaining background in future DSNB searches in SK-Gd.
The dominant background in the high-energy region comes from electrons or positrons via atmospheric neutrino interactions,
while the dominant background in the low-energy region is the decay of radioactive isotopes, especially $^9\rm{Li}$, produced by the spallation of oxygen nuclei by cosmic ray muons.
The former is expected to lead to a better understanding of the neutron multiplicity of the final state of atmospheric neutrino interactions, which is one of the major uncertainties, by constructing a more precise nuclear interaction model.
The latter is expected to be further reduced by more precise analysis using a simulation-driven method for the spallation background estimation instead of the current data-driven method.

\subsection{Neutrons from Atmospheric Neutrinos}
The increased neutron detection efficiency has benefits for atmospheric neutrino observation as described in Section~\ref{sec:atmnu-bene}.
Since the start of SK-Gd, we have published two papers on atmospheric neutrino interactions~\cite{sakai:2024,Han:2025}.
Atmospheric neutrinos are observed by measuring the particles produced when they interact with oxygen nuclei in water.
The improvement in neutron detection efficiency at SK-Gd has provided important insights, particularly into hadron interactions in water using the neutrons produced by neutrino interactions.
Neutrino interactions at the GeV scale, such as those observed with atmospheric neutrinos, are modeled in three steps as follows:  sampling particles emitted from the initial neutrino interaction at the nucleon or quark level, intranuclear hadron transport and subsequent nuclear de-excitation, and hadron transport and nuclear deexcitation in water.

The first paper~\cite{sakai:2024} focused on events observed at SK with particle energies below 30~MeV, called neutral-current quasi-elastic [NCQE] events, which can be expressed as
\begin{eqnarray}
    \nu(\bar{\nu}) + ^{16}\rm{O} \rightarrow \nu(\bar{\nu}) + ^{15}\rm{O} + \gamma + \rm{n}, \nonumber \\
    \nu(\bar{\nu}) + ^{16}\rm{O} \rightarrow \nu(\bar{\nu}) + ^{15}\rm{N} + \gamma + \rm{p}. \nonumber
\end{eqnarray}
When gamma rays and a single neutron are observed resulting from this interaction, it becomes a serious background in the DSNB search because the reconstructed gamma ray energy is similar to the positron energy in IBD from DSNB, and it is difficult to distinguish this background from the DSNB signal.
Therefore, the NCQE signal itself was investigated. 
In this study, several simulations were compared with data regarding the PMT hit pattern and reconstructed energy of the prompt gamma ray signal, and the neutron multiplicity obtained from the delayed signal.
The major difference observed here was the hit pattern from the prompt gamma rays, and the result showed that the data for prompt gamma ray multiplicity was significantly less than in the simulation previously used.
Further investigation revealed that the cause of this difference was nuclear de-excitation, which could be improved by using the Precompound model to simulate the emission of baryons from recently struck, highly excited nuclei, as implemented in Geant4~\cite{precompound:2011}.
This is consistent with the results of an external (non-SK) experiment in which a water target was irradiated with a neutron beam and the generated gamma rays were measured with a germanium detector.

In the second paper~\cite{Han:2025}, we report the results of an atmospheric neutrino analysis in the energy range from 30~MeV to 10~GeV, which covers almost the entire observational range at SK.
This analysis focused particularly on neutron production versus energy of the prompt event, and systematically compared the data with various simulations, including models for all the processes described above.
Here, the predicted neutron production differed by up to 50\% from the simulation, demonstrating strong model discrimination power.
Based on this analysis, two key results were obtained.
First, neutron production is reduced in sub-GeV single-ring events, where the contribution of secondary hadron interactions is small.
This result highlights the need to consider appropriate neutron evaporation and nuclear in-medium effects in nuclear reaction models.
Second, neutron production increases almost linearly with the energy of prompt events, especially in the multi-GeV region.
This result is more strongly influenced by pion production in the cascade model, which becomes important at higher energies.
In conclusion, we found that the de-excitation model previously used in the hadron-nucleus interaction model in nuclear de-excitation is inconsistent with the observational data; this discrepancy between the de-excitation model and data was also seen in the first paper.
Our data also support several specific models that reduce the uncertainty in total neutron production from atmospheric neutrino events to about 10\%.

Finally, these results demonstrate that increased neutron capture efficiency has led to improved nuclear reaction models, which in turn can be expected to contribute to improvement for all neutron-related data analysis at SK, such as the identification of atmospheric antineutrino events, enhancing sensitivity to neutrino mass ordering, and searches for proton decays and the DSNB.

\section{Prospects for Future DSNB Detection with SK-Gd}
With the addition of gadolinium to the Super-Kamiokande detector, the sensitivity to the DSNB electron antineutrino ($\bar{\nu}_e$) flux has significantly improved, particularly due to enhanced neutron tagging efficiency. However, the current DSNB search analysis is still limited by both systematic and statistical uncertainties, particularly in the estimation of atmospheric neutrino backgrounds such as NCQE and non-NCQE interactions.

We have adopted a data-driven approach for the non-NCQE normalization uncertainty, using the high-energy sideband just above $E_{\rm rec} = 29.5$ MeV. While the limited statistics currently constrain this method, these uncertainties are expected to shrink with continued data accumulation. Indeed, the SK-IV period showed roughly half the systematic uncertainty seen in SK-VI and SK-VII, suggesting that similar improvements are likely for upcoming SK-Gd data.

Further reductions in systematic uncertainties are anticipated from advances in modeling atmospheric neutrino interactions, especially those involving oxygen nuclei. Recent theoretical work, such as \cite{zou:2024}, has shown encouraging agreement with our atmospheric-neutrino data. Moreover, recent experimental efforts by Super-K~\cite{sakai:2024}, as well as the study by T2K~\cite{Licheng:2025}, have begun to refine our understanding of the neutrino-oxygen NCQE cross sections and the neutron multiplicity in the final state. Ongoing validation of these models with data is crucial, particularly for better characterizing high-multiplicity neutron events that can mimic DSNB signals.

At low energies, ${}^9$Li spallation remains the dominant background. Lowering the energy threshold for DSNB searches is particularly important for spectral fits, and probing the predicted low-energy region of DSNB models. Continued improvements in spallation rejection techniques will be essential to extending sensitivity below 13.5 MeV.

\begin{figure}[htbp] 
\includegraphics[width=0.8\textwidth]{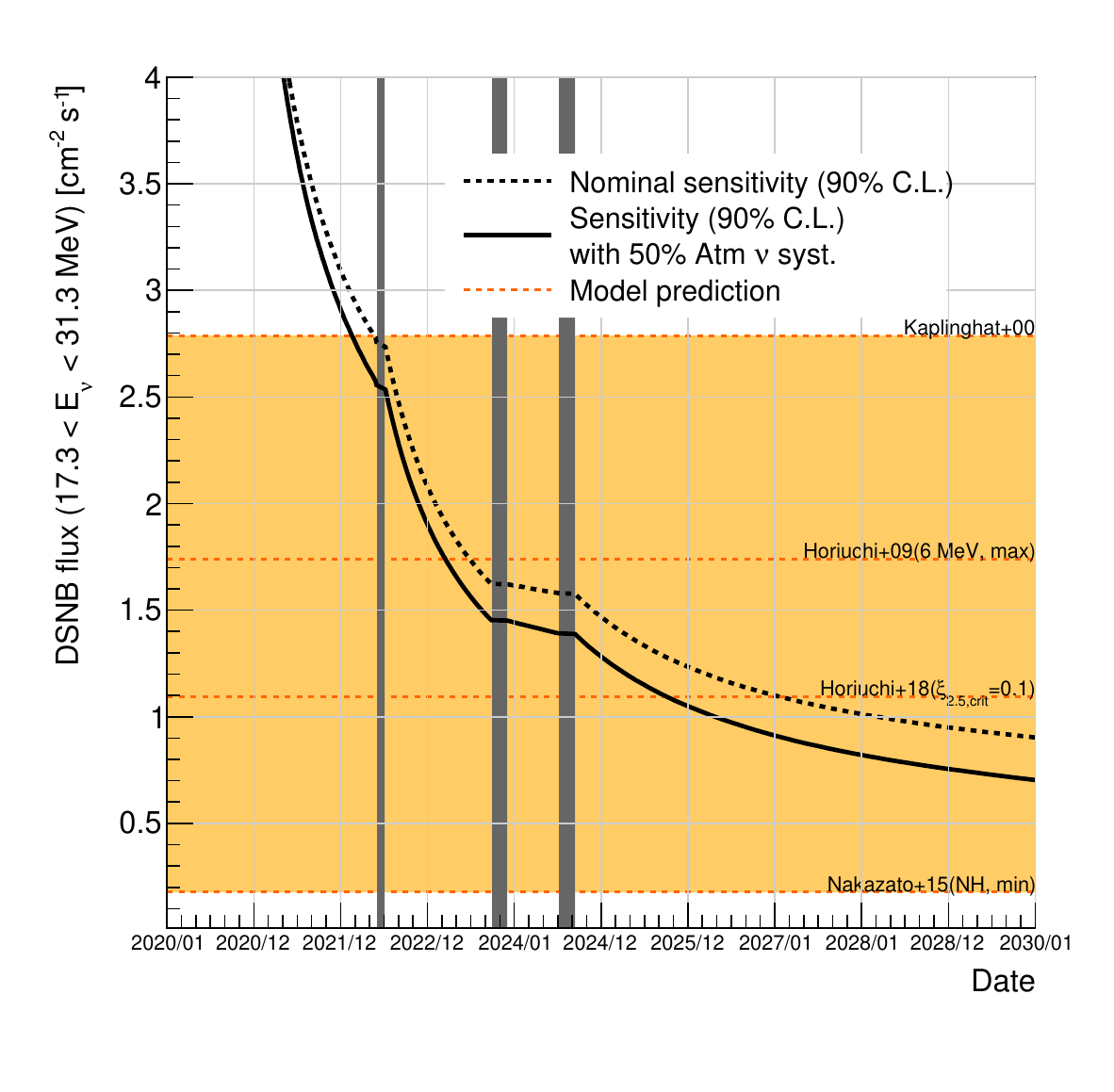} 
\caption{Projected 90\% confidence level sensitivity to the DSNB flux in SK-Gd as a function of time. The dotted curve assumes current background estimates, while the solid curve assumes a 50\% reduction in the atmospheric neutrino background. The shaded band represents the range of theoretical DSNB model predictions. The temporary sensitivity degradation in 2024 due to a failure of the geomagnetic compensation coil for the PMTs, and the subsequent repair, is taken into account.  Vertical bars indicate periods during which useful DSNB data could not be taken by SK due to unstable detector conditions. } \label{fig:skgd_sensitivity} 
\end{figure}

Figure~\ref{fig:skgd_sensitivity} shows the projected 90\% confidence level sensitivity of SK-Gd to the DSNB $\bar{\nu}_e$ flux as a function of time, under both current background estimates and a scenario in which the atmospheric neutrino component is reduced by 50\%. These projections demonstrate that SK-Gd will soon explore a substantial region of the parameter space predicted by leading DSNB models, including those by Horiuchi et al.~\cite{Horiuchi:2009, Horiuchi:2018}, Kaplinghat et al.~\cite{Kaplinghat:2000}, and Nakazato et al.~\cite{Nakazato:2015}.

In the higher energy regime, the DSNB flux prediction is particularly sensitive to contributions from failed supernovae (i.e., core-collapse events that form black holes), since such progenitors emit more energetic neutrinos.  A robust detection of this high-energy tail would provide unique insight into the cosmic history of stellar death and black hole formation.


\section{Conclusion}
The gadolinium upgrade of Super-Kamiokande represents a major milestone in neutrino experiments. Years of R\&D, including the EGADS project, material purification, detector refurbishment, and careful implementation of gadolinium loading, have culminated in a fully operational water Cherenkov detector with neutron-tagging capability. The initial operation of SK-Gd has confirmed the expected performance: gadolinium is homogeneously distributed, radiopurity goals have been met, and neutron detection efficiency has been significantly improved. These achievements enable unprecedented sensitivity to the diffuse supernova neutrino background, bringing the long-sought detection of this signal within reach.
In addition, SK-Gd improves the ability of the experiment to study galactic supernovae, atmospheric neutrino interactions, reactor neutrinos, proton decay, and pre-supernova neutrinos, thus broadening the physics program of Super-Kamiokande. The techniques developed for SK-Gd also provide valuable guidance for future large-scale detectors, including Hyper-Kamiokande. With the SK-Gd upgrade, the field enters a new era in which the detection of the DSNB, and the rich astrophysical insights it carries, appear to be increasingly imminent.

\section*{DISCLOSURE STATEMENT}
The authors are not aware of any affiliations, memberships, funding, or financial holdings that
might be perceived as affecting the objectivity of this review. 

\section*{ACKNOWLEDGMENTS}
We gratefully acknowledge cooperation of the Kamioka Mining and Smelting Company.
The Super-Kamiokande experiment was built and has been operated with funding from the
Japanese Ministry of Education, Science, Sports and Culture, 
and the U.S. Department of Energy.
This work was supported by JSPS KAKENHI Grant Numbers JP17H06365, JP19H05807, JP20H00162, JP21224004, JP26000003, JP21H04989, JP24H00243, and JP24H02242.

%


\end{document}